\documentclass[sigconf]{acmart}
\usepackage{booktabs}
\usepackage{tikz,xcolor} 
\usepackage{amsmath,graphicx}
\usetikzlibrary{matrix, positioning}
\tikzset{main node/.style={circle,fill=white,draw,minimum size=1cm,inner sep=0pt},} 
\usetikzlibrary{decorations.pathreplacing}
\tikzset{main node/.style={circle,fill=white,draw,minimum size=3cm,inner sep=1pt},}
\tikzset{
	place/.style={
		circle,
		thick,
		draw=black,
		fill=gray!50,
		minimum size=6mm,
	},
	state/.style={
		circle,
		thick,
		draw=blue!75,
		fill=blue!20,
		minimum size=6mm,
	},
}
\graphicspath{ {images/} }
\usepackage{txfonts}
\usepackage{pxfonts}
\usepackage{tabularx}
\AtBeginDocument{%
  \providecommand\BibTeX{{%
    \normalfont B\kern-0.5em{\scshape i\kern-0.25em b}\kern-0.8em\TeX}}}

\copyrightyear{2021} 
\acmYear{2021} 
\setcopyright{acmcopyright}\acmConference[FAccT '21]{Conference on Fairness, Accountability, and Transparency}{March 3--10, 2021}{Virtual Event, Canada}
\acmBooktitle{Conference on Fairness, Accountability, and Transparency (FAccT '21), March 3--10, 2021, Virtual Event, Canada}
\acmPrice{15.00}
\acmDOI{10.1145/3442188.3445886}
\acmISBN{978-1-4503-8309-7/21/03}

\begin{document}

\title{The Use and Misuse of Counterfactuals in Ethical Machine Learning}

\author{Atoosa Kasirzadeh}

\affiliation{%
  \institution{University of Toronto  \and Australian National University}
}
\email{atoosa.kasirzadeh@anu.edu.au}

\author{Andrew Smart}
\affiliation{%
 \institution{Google}
}
\email{andrewsmart@google.com}

\begin{abstract}
The use of counterfactuals for considerations of algorithmic fairness and explainability is gaining prominence within the machine learning community and industry. This paper argues for more caution with the use of counterfactuals when the facts to be considered are social categories such as race or gender. We review a broad body of papers from philosophy and social sciences on social ontology and the semantics of counterfactuals, and we conclude that the counterfactual approach in machine learning fairness and social explainability can require an incoherent theory of what social categories are. Our findings suggest that most often the social categories may not admit counterfactual manipulation, and hence may not appropriately satisfy the demands for evaluating the truth or falsity of counterfactuals. This is important because the widespread use of counterfactuals in machine learning can lead to misleading results when applied in high-stakes domains. Accordingly, we argue that even though counterfactuals play an essential part in some causal inferences, their use for questions of algorithmic fairness and social explanations can create more problems than they resolve. Our positive result is a set of tenets about using counterfactuals for fairness and explanations in machine learning.
\end{abstract}



\acmYear{2021}\copyrightyear{2021}
\setcopyright{acmcopyright}
\acmConference[FAccT '21]{ACM Conference on Fairness, Accountability, and Transparency}{March 3--10, 2021}{Virtual Event, Canada}
\acmBooktitle{ACM Conference on Fairness, Accountability, and Transparency (FAccT '21), March 3--10, 2021, Virtual Event, Canada}
\acmPrice{15.00}
\acmDOI{10.1145/3442188.3445886}
\acmISBN{978-1-4503-8309-7/21/03}

\maketitle

\section{Introduction}

The use of counterfactuals has become increasingly popular in the machine learning community for many reasons such as making sense of algorithmic fairness or explainability in automated decision-making for consequential social contexts \cite{kusner2017counterfactual,wachter2017counterfactual,barocas2020hidden,coston2020counterfactual,ghai2020measuring,mothilal2020explaining,hendricks2018generating,sokol2018glass,grath2018interpretable}. As a result, machine learning algorithms coupled with counterfactuals could be used for making high-stakes decisions with ethical and legal impacts in domains such as insurance, predictive policing, and hiring. Despite this widespread attention and use, there is a surprising lack of engagement with the long-standing philosophical and social scientific literature on the required ontological and semantic conditions for an appropriate application of counterfactuals. 

What is a counterfactual? Consider X and Y to represent events or facts and the following chain of occurrences ``X and Y'', where X precedes Y in time. A counterfactual analysis can help to find whether X is a cause of Y by supposing the non-occurrence of X and seeking for the effect of this supposition on Y. This corresponds to evaluating whether the counterfactual `If X had not occurred, Y would not have occurred.' is true. In machine learning practice, there are several technical ways to generate and evaluate counterfactuals, such as feature-based explanations, prototype explanations, example-based explanations, or causal explanations \cite{kusner2017counterfactual,wachter2017counterfactual,ribeiro2016should,lundberg2017unified,guidotti2018local,van2019interpretable,joshi2019towards}. These approaches are most often rooted, implicitly or explicitly, in either of the two prominent conceptual approaches for evaluating counterfactuals: the close-enough-possible-worlds approach inspired by Lewis \cite{lewis1973counterfactuals} and Stalnaker \cite{stalnaker1968theory}, and the causal modeling approach developed by Spirtes et al. \cite{spirtes2000causation} and Pearl \cite{pearl2009causality}, among others.\footnote{Strictly speaking, \cite{stalnaker1968theory,lewis1973counterfactuals} develop the closest-possible-worlds approach to make sense of counterfactuals. With a bit of weakening, the (set of) closest-possible-world(s) can be interpreted as the (set of) close-enough-possible-world(s), where due to practical considerations those possible worlds that are close enough to the actual world (rather than the closest possible worlds) are selected. For a recent alternative to evaluating conditionals relative to a causal model see \cite{andreas2020ramsey}.} 

To evaluate a counterfactual, the close-enough-possible-worlds approach compares the actual world in which X and Y occur with those similar-enough worlds to the actual world in which X does not occur (e.g., comparing a data instance to a similar data instance or to a prototype when generating example-based or prototype explanations, respectively, requires comparison with respect to some notion of enough similarity). If in those worlds Y does not occur the counterfactual is considered true and X is deemed the cause of Y; otherwise, the counterfactual is deemed false. The close-enough-possible-worlds account has been mainly used in discussions of counterfactual explanations in machine learning and the causal modeling approach has been widely applied for examining fairness counterfactually. Although these two semantic accounts are very different, the following abstract recipe is common to both for the evaluation of counterfactuals. First, determine the facts to be kept fixed under counterfactual variation. Second, vary the antecedent. Third, determine the influence of the variation on the consequent.

In this paper, we explore the ontological and epistemological-semantic conditions required for using either of the two conceptual approaches for an appropriate application of counterfactuals to ethical machine learning, in particular to algorithmic fairness and social explanations. We argue that in some cases, the lack of a right grounding of the elements of a counterfactual into the social world can lead to their misuse in machine learning applications. We review a broad body of papers from philosophy and social sciences on the ontology of social categories and conclude that the counterfactual approach in machine learning fairness and social explainability might require an incoherent theory of what some social categories such as race are. Our findings suggest that despite its appeal for convenient analysis of fairness and social explanations, most often the social categories may not admit an apt counterfactual intervention, and hence may not appropriately satisfy the required assumptions for evaluating the truth or falsity of counterfactuals. Accordingly, we argue that even though counterfactuals play an essential part in some causal inferences, their use in discussions of algorithmic fairness and social explanations can create more problems than they resolve.

\textbf{Related work and novelty.} Before we go further, we would like to explicitly contrast our paper in more detail with related work to highlight its novelty. There are four main closely related works on this topic which explicitly or implicitly critique counterfactual theories of social causation in decision-making contexts. Kohler-Hausmann \cite{kohler2018eddie} argues that the counterfactual causal model is \emph{wrong} for detecting discrimination in both law and social science. Building on this idea, Hu and Kohler-Hausmann \cite{hu2020s} argue that perhaps we need to use a formal model other than causal models (such as constitutive diagrams) for detecting discrimination. Hanna et al. \cite{hanna2020towards} use critical race theory and argue that the multi-dimensionality of race should be taken into account whenever this phenomenon becomes relevant to the machine learning community, and challenges practitioners to explicitly ask who is
doing the categorizing and for what purpose? Barocas et al. \cite{barocas2020hidden} discuss the mapping of the explanatory features to actions in the world when using feature-highlighting explanations. We share the perspective of these authors. However, the novelty of our contribution is threefold. (1) We provide a conceptual analysis of the vagueness of the notion of ‘similarity’, rooted in the close-enough-possible-worlds approach. This approach is the conceptual basis of feature-based, prototype, and example-based analytic methods for examining counterfactuals by machine learning community. The notion of similarity is used in almost all conceptions of counterfactual explanations or fairness as referenced. To the best of our knowledge, the philosophical-conceptual basis \cite{stalnaker1968theory,lewis1973counterfactuals,lewis1986philosophical} and assumptions required to assess the ‘similarity’ of counterfactual worlds/scenarios are not properly examined in the machine learning literature, yet ‘similarity’ is used, implicitly or explicitly, for making sense of counterfactual explanations or fairness. (2) We go beyond the mere criticism of causal modeling as applied to the social domain, and consider counterfactuals more generally by examining both the close-enough-possible-worlds account and causal modeling. We think that just a critique of manipulating social categories is not sufficient because in disciplines such as medicine and public health, the use of protected attributes such as race or gender are considered to be an ethically acceptable component of research (e.g., prostate cancer screening \cite{ben2008risk,halabi2019overall}).\footnote{We are not promoting this use. We just report that in medicine, economics, public health and other related disciplines, the use of protected classes such as race or gender sometimes is the basis of development or allocation of some resources.} (3) We provide positive results in terms of a set of detailed tenets as summarized in table 1, showing that any trace of a counterfactually fair or explainable algorithm (in a social context) involves making several choices and value judgments. To that end, the implicit presumptions, choices, and value judgments must be made as explicit and obvious as possible by using table 1. No related work does (1) -- (3).

The rest of the paper is structured as follows. In Section 2, we examine the two prominent approaches to modeling and evaluating counterfactuals, the close-enough-possible-worlds and the causal modeling approaches, in more detail. In Section 3, we discuss the use of counterfactuals for analyzing fairness and social explanations in machine learning practice before raising ontological and epistemological-semantic problems from this use in Section 4. In Section 5, we suggest a set of tenets about the use of counterfactuals in machine learning. Section 6 concludes the paper.

\section{Background: the close-enough-possible-worlds and causal modeling}

Consider the following counterfactuals: (1) If Suzy had not thrown the rock, the window would not have shattered. (2) If Nora had not been Latina, she would not have been denied admission. Are these counterfactuals true or false? Does `Suzy's throwing the rock' cause `the shattering of the window'? Does `Nora's being Latina' cause `denying admission'? There are two prominent approaches to evaluate counterfactuals, the close-enough-possible-worlds approach that is mainly used in the discussions of social counterfactual explanations \cite{wachter2017counterfactual}, and the causal modeling approach that is at the center of discussions about counterfactual fairness \cite{kusner2017counterfactual}.\footnote{Kilbertus et al. \cite{kilbertus2017avoiding} use causal models to analyze fairness. We focus our discussion on Kusner et al. \cite{kusner2017counterfactual}, but our criticism also applies to their work.}  We present these two semantic approaches independently, though we must mention that, theoretically speaking, the relationship between the two is not that straightforward \cite{briggs2012interventionist}. For the lack of space, we cannot go into the differential details in this paper. But we translate this lack of straightforward connection between the two semantic approaches into our set of principles for using counterfactuals in machine learning research.

According to the closest-possible-worlds view \cite{stalnaker1968theory,lewis1973counterfactuals}, a counterfactual can be treated syntactically and semantically via a variant of a modal logic for counterfactuals. The evaluation of the counterfactual X $\boxright$ Y (if X had occurred, Y would have occurred) requires the specification of a set of possible worlds in which X occurs. If in these possible worlds Y also occurs, the counterfactual X $\boxright$ Y is true. These possible worlds must be ordered in terms of comparative similarity or closeness to the actual world (in which X occurs and Y occurs). For instance, if in all the worlds which are close enough to the actual world except that Suzy does not throw the rock, the window does not shatter, then Suzy's throw is the cause of the shattering of the window. If in all the close-enough-possible-worlds to the actual world in which Nora is not Latina, she is not denied admission, then Nora's being Latina is the cause of her rejection. The close-enough-possible-worlds approach to the evaluation of counterfactuals requires an ordering of the possible worlds in terms of similarity to the actual world. In Section 4, we discuss that the notion of \emph{similarity} is inherently vague and that the similarity ordering can be done in many different ways. As a result, depending on \emph{the choices} for the similarity criteria and the ordering, we can obtain contradictory judgments about the truth or falsity of counterfactuals. Hence, the vagueness and the multiplicity of orderings pertain to the problems of using counterfactuals in machine learning. 

A causal modeling approach uses a causal model as a representational tool for exploring the space of alternative causal hypothesis. Following Pearl \cite{pearl2009causality}, from a causal modeling perspective, the world is described in terms of random variables and their values. The random variables are either exogenous or endogenous, and they might take continuous or categorical values. The exogenous variables ($\mathcal{U}$) are determined by factors outside of the causal model, and serve as fixed background assumptions to the causal reasoning. The endogenous variables ($\mathcal{V}$) may have a causal influence on each other. This influence is modeled by a set of structural equations $(\mathcal{F})$ that are functions for capturing the potential causal effects of functional dependencies on the endogenous variables. A set of exogenous and endogenous variables, their values, and a set of structural equations form a causal model $\mathcal{M}$=($\mathcal{U},\mathcal{V},\mathcal{F}$). $\mathcal{M}$ can be graphically visualized by a directed acyclic graph. This graph facilitates cognitive efforts in thinking about potential causal sources, effects, and causal relations. In such a graph, a node represents a random variable and an edge between each pair of nodes represents a direct causal relation between the corresponding random variables; for instance, $X$ is a direct cause (parent) of $Y$ is represented by $X\rightarrow Y$. Nodes with no incoming edge are said to be exogenous.

To find causal relations via a causal model requires establishing well-defined connections between some aspects of the sample data and a causal model \cite{spirtes2000causation,pearl2009causality}. The main connections are often captured by two causal assumptions, the causal Markov condition and faithfulness. The causal Markov condition ensures that a variable is independent of its non-descendants given its parents. The causal faithfulness condition requires that all inter-dependencies in the observational data are non-accidental and structural, the result of the structure of the causal graph. To counterfactually think via a causal modeling approach in a specific machine learning domain requires an in-depth interpretation of the mapping of the random variables on the elements of the domain and the satisfaction of the causal assumptions. If the domain of counterfactual thinking occurs at the level of the social world, we require an apt interpretation of the mapping of the random variables on social categories, the relationship between them, and the meaning of causal assumptions applied to the relevant categories. So far, we have provided a discussion of the two most prominent semantic approaches to the evaluation of counterfactuals. In the next section, we give two examples of the use of counterfactuals in machine learning: in understanding fairness (via causal modeling) \cite{kusner2017counterfactual} and in understanding social explanations (via the closest-possible-worlds) \cite{wachter2017counterfactual}.

\section{Counterfactuals in ethical machine learning}

\subsection{Counterfactual fairness}

Discussions about the treatment of \emph{fairness} in machine learning systems have primarily taken place in relation to a group or the individual level. To achieve group fairness, a (statistical) measure must compare a predictor's behavior across different protected demographic groups, and then seeks for approximate parity of some desirable statistical measure across the groups \cite{hardt2016equality,chouldechova2017fair}. On the other hand, a measure of individual fairness must compare a predictor's behavior across similar individuals \cite{dwork2012fairness,joseph2016fairness}. To date, the most popular proposal for making sense of individual fairness has been the use of causal modeling for interpreting individual fairness in a counterfactual way \cite{kusner2017counterfactual}. Kusner et al. \cite{kusner2017counterfactual} define a fair predictor to be the one that gives the same prediction had the individual were different, for example, had the individual been of another race or gender. This demands an implicit assumption that other features and properties (except for the tweaked category in the causal model) remain the same for that individual. More precisely, Kusner et al. (2017) gives the following definition: counterfactual fairness ``captures the intuition that a decision is fair towards an individual if it is the same in (a) the actual world and (b) a counterfactual world where the individual belonged to a different demographic group.''

Consider a prediction-based problem characterized in terms of $A$ (a set of protected attributes), $X$ (a set of non-protected attributes), and $Y$ (the prediction output). To put this problem into a causal modeling schema requires fixing $U$, the set of exogenous variables. Following \cite{kusner2017counterfactual}, the definition of counterfactual fairness for the predictor $\hat{Y}$ stipulates the satisfaction of the following condition for $X$=$x$ and $A$=$a$, for all $y$, and any value $a^\prime$ attainable by $A$:

$P$($\hat{Y}_{A\leftarrow{a}}$($U$)=$y|X$=$x,A$=$a$)=$P$($\hat{Y}_{A\leftarrow{a^\prime}}$($U$)=$y|X$=$x,A$=$a^\prime$)

To make matters more concrete, we focus on an example of a machine learning system, as discussed by \cite{kusner2017counterfactual}, employing a predictor $\hat{Y}$ to decide who should be admitted to law school based on its prediction of potential student's first year grade (figure 1). The algorithm makes the prediction according to knowledge about the following attributes of individuals: gender, race, GPA, and law school entrance exam  (LSAT). According to \cite{kusner2017counterfactual}, the set of sensitive attributes are $A=$\{sex, race\}, and the non-sensitive ones are $X=$\{GPA, law school entrance exam\}. Moreover, there is a causal link set between the attributes and the prediction of potential student's first year grade. To make this classifier fair, the following question should be answered: what would the predictor have predicted, if the individual had a different race (a different sensitive attribute)? This use of counterfactuals requires assuming a single change (race) or a limited set of changes (such as sex and race) to an individual, and then evaluate the probabilistic condition above given the supposition that everything else remains the same for that individual. Although the proposal might sound simple, in the next section we discuss the problems pertaining to this proposal such as requiring commitment to a peculiar conception of race as well as controversial views about the integrity of what an individual (or the perception of an individual) is, for the purpose of satisfying the convenient requirements of counterfactual modeling and evaluation. Or, to understand what counterfactual fairness is, we first need to make choices about which counterfactual worlds to consider and the basis by which the closeness of counterfactual worlds (including the knowledge of how a counterfactually different version of the target individuals) to the actual world is specified.

\begin{figure}
\centering
\resizebox{4cm}{3cm}{
\begin{tikzpicture}[node distance=1cm ,auto, every place/.style={draw}]

%
\node [place] (1) {\small $Race$};
\node [place] (2) [above right  =of 1] {\small $ GPA$};
\node [place] (3) [below =of 2] {\small $LSAT$};
\node [place] (4) [below =of 3] {\small $Grades$};

\draw[->, thick] (1) -- node [sloped,above] {} (2);
\draw[->, thick] (1) -- node [sloped,above] {} (3);
\draw[->, thick] (1) -- node [sloped,above] {} (4);

\node [place] (5) [below =of 1] {\small Sex};
\draw[->, thick] (5) -- node [sloped,above] {} (2);
\draw[->, thick] (5) -- node [sloped,above] {} (3);
\draw[->, thick] (5) -- node [sloped,above] {} (4);

\node [place] (6) [right =of 3] {\small $Know$};
\draw[->, thick] (6) -- node [sloped,above] {} (2);
\draw[->, thick] (6) -- node [sloped,above] {} (3);
\draw[->, thick] (6) -- node [sloped,above] {} (4);
\end{tikzpicture}
}
\caption{A causal model for a fair predictor adapted from \cite{kusner2017counterfactual}.}
\end{figure}
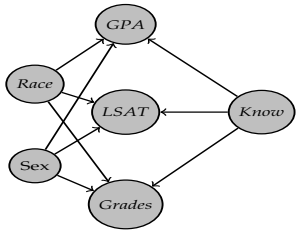

\subsection{Social counterfactual explanation}

Counterfactual explanations are claimed to be among the most popular types of explanations for opaque algorithmic decisions \cite{wachter2017counterfactual}. For instance, let us assume Nora has applied for a mortgage and her application is denied via an algorithmic system. A counterfactual explanation for this denial can be: If Nora's annual income had been \$60,000, she would have received the loan. As a matter of fact, Nora is denied a loan and her annual income is \$40,000. Or consider the following counterfactual explanation: If Nora had not been Latina, she would not have been denied the loan. As a matter of fact, Nora is denied a loan and she is Latina. This is an instance that requires making a putative plausible assumption about a different version of Nora (with only a different race, everything else equal to the original version of Nora), and then trust the validity of this explanation.

In the next section, we offer two main arguments challenging a counterfactual approach to algorithmic fairness and social explanations when the things that require counterfactual supposition are social categories such as gender, sexual orientation, or race in terms of which (the uniqueness of) a person is characterized. 

\section{Two problems}

In this section, we specify two sets of problems, ontological and epistemological-semantic, that one faces upon attempting to construct a fair or explainable classifier which incorporates the counterfactual supposition of social categories such as race and gender, as sketched in Section 3. The problems arise in the attempts to answer the following questions: what are the objects of manipulation? Which counterfactual worlds are similar enough to the actual world or whose causal model's perspective should we care about?

\subsection{What is manipulated?}

There has been a long standing debate among several disciplines such as philosophy, sociology, law and epidemiology about the causal effects of social categories such as race and gender \cite{kohler2018eddie,elder2012towards,glymour2014commentary,marcellesi2013race}. To counterfactually suppose a social attribute \emph{of an individual} requires first specifying what the social categories are and what it means to suppose a different version of an individual with the counterfactually manipulated social property. This counterfactual question amounts to asking, ''what if person X had not been ''race Y'' or ''gender Z''? 

There are several competing contemporary schools of thought about what social categories are, and our review here is merely representative of some and by no means exhaustive. In the rest of this section, we take `race' as a prototypical instance of the social categories of interest to counterfactual manipulation. With some modifications, similar arguments can be made about other social categories such as gender. 

Roughly, we can distinguish between three major positions about what race is \cite{mallon2004passing,mallon2007field}. The geo-biological essentialism about race largely signifies dividing humans into a sufficiently small, discrete number of categories, usually for the purposes of colonial conquest, enslavement or domination of one group over another \cite{dance2010struggles}. The categorization has been based on some kind of biological foundation (e.g., modern genes) essential to humans, and inherited from one generation to another. This conception of race identifies some geo-biological features (such as skin color, hair texture, and eye form) that are only common to the members of a racial group, usually from a specific geographical region. The geo-biological conception of race has been questioned extensively, and has been critically challenged by scientific and philosophical arguments ranging from denying that the concept of race has any biological foundations to denying the very existence of races. Some also have argued that this biologically essentialist view about race cannot be separated from the political project of racial oppression, domination and disenfranchisement \cite{dance2010struggles,hanna2020towards}. In addition to the geo-biological ancestry conception of race, there are two other major views about the ontology of race.


On the one hand, racial skeptics argue for the falsity of naturalism about race and conclude that no type of race exists \cite{appiah1995uncompleted,appiah1996race,zack1994race,hacking1999social,zack2014philosophy,wagner2017anthropologists}. They claim that the natural candidates for the bases of race such as geography, phenotypes, and geneology fail according to scientific findings. The normative implications of this ontological view is to entirely disregard the existence of race. On the other hand, racial constructivists dismiss the conception of biological race, but argue that the concept of race must be preserved for the purpose of social movements and affirmative action to abolish social and structural injustice. How so? One of the most influential proponents of racial constructivism, Haslanger \cite{haslanger2000gender,haslanger2010language,glasgow2019race}, suggests a group-based understanding of race marked by ancestry and appearance and by hierarchical relations of power \emph{for the purposes of fighting against social injustice}. This conception of race finds using `race' as a justifiable entity for the purpose of resisting and combating racism. Other than that, racial identification by the dominant group constrains the autonomy of individuals by requiring them to be what a specific racial group signifies from the point of view of who has defined it. Social constructivism hence maintains that a social category -- be it racial, gender, or class – was brought into existence or shaped by historical events, social forces, political power, or colonial conquest, all of which could have been very different \cite{hacking1999social,elder2012towards,benjamin2019race}. Being a social constructivist about race and gender means that one does not subscribe to the view that race and gender are natural or biological categories with permanent or immutable properties. In other words, for such a constructivist, the term `race' cannot refer to an essentially biological attribute such as skin tone, a genetically produced trait, or a signifier that people just have and thereby obviously belong to a designated racial group \cite{kohler2018eddie}. 

Kusner et al. \cite{kusner2017counterfactual} claim that it is counterproductive to assume social categories such as race cannot be causes because we can design experiments on such categories by intervening on a particular aspect of the attribute `race', such as `race perception'. We disagree. We think this claim only serves to justify the convenient assumptions required for causal modeling (i.e., that conception of race is amenable to counterfactual manipulation). As we have shown above, there is no universally agreed-upon perception of race. To be able to talk about the causal effect of social categories, we first need to specify what these categories are. For instance, we might be justified in first having a robust social ontology informed by critical theory \cite{haslanger2016social}. Only after this exploration, we are able to discuss what our perception of race is. As we have seen, there is a plurality of responses to this question, and our response depends on the perspective we adopt about this matter.

Recall that an algorithm that subscribes to counterfactual fairness requires evaluating the actual non-occurrence of X with the supposition that X did occur. For example, we should replace the actual person (or our perception thereof) who has a protected attribute, such as being Latina, with a counterfactual version of the same person who has a different protected attribute, such as being white, to test whether the algorithm makes the same prediction about the actual person (or our perception thereof) and the counterfactual person. What view about race (or perception of race) does it require to suppose that racial category non-Latina for the counterfactual version of person $i$ knowing that Latina is the real feature of person $i$? Counterfactual fairness (or counterfactual social explanation) requires us to force a random variable to take a certain value. Is the required counterfactual suppositions for designing a fair algorithm compatible with the views about race specified above? 

Racial skepticism is ruled out as an alternative of commitments held by the proponents of counterfactual fairness or counterfactual social explanations due to its denial of the very existence of such categories. Social constructivism makes sense of race \emph{for the purposes of fighting against social injustice}. Hence, the constructivist ontology of race has, in addition, a purpose-relative reality that the algorithm must reflect in its reasoning and arguably is not subject to counterfactual variation separate from the scope of the fight against social injustice. Perhaps the only viable theory of race that remains for counterfactual fairness requires commitment to a reductionist view about social categories such as race or gender as biological attributes. Several scholars have argued that this commitment is deeply problematic (see, for instance, \cite{kohler2018eddie}). We share this perspective for several decision contexts. This purely reductionist understanding of social categories as essential and physical attributes, in addition to being scientifically outdated, fails the task of robust objectivity, and might indirectly widen and exaggerate the problematic associations between the sensitive attributes that are the result of social and structural injustice in the first place. 

\subsection{Similarity between worlds and the view from somewhere}

Is there an objective view from nowhere form which to assess the validity of counterfactuals? In this section, we raise some epistemological-semantic problems for comparing and selecting the set of counterfactual possible worlds that are close enough to the actual world. 

First, we focus on the problem of inherent vagueness associated with similarity between possible worlds. Counterfactual scenarios in counterfactual worlds stand in contrast to actual scenarios in actual worlds. To evaluate a counterfactual requires a comparison between an actual world and a set of sufficiently similar counterfactual worlds to the actual world. The counterfactual $X \boxright Y$ is true just in case it takes less of a departure from the actual world to make $X$ true along with $Y$ than to make $X$ true without $Y$. But, which counterfactual worlds? The number of counterfactual worlds is myriad (perhaps even uncountably infinite). Lewis and Stalnaker \cite{stalnaker1968theory,lewis1973counterfactuals} emphasize that the counterfactual worlds of interest to the actual world are the ones that are the most similar to the actual world. In some cases of comparing natural features between worlds, it is possible to arrive at a consensus for the ordering of similar worlds. However, in many cases the vagueness of this notion is problematic and counterintuitive for the evaluation of counterfactuals \cite{goodman1972seven}. Lewis \cite{lewis1986philosophical} provides some guidance to ordering possible worlds: (1) avoid big widespread violations of the laws of nature of the actual world, (2) maximize the spatiotemporal perfect match of particular matters of fact, (3) avoid small, localized violations of the laws of nature of the actual world, and (4) secure approximate similarity of particular matters of fact. But, how to translate these considerations to the social domain? Further research is required to understand how to avoid big widespread violations of commitments to our ontological views about social categories in the possible worlds framework. 

The ordering of similar worlds faces severe problems because for some ordinary counterfactuals, some irrelevant possible worlds end up determining the counterfactuals' truth values. Also, depending on what kind of possible worlds we choose, we might end up assigning a different truth-value to a counterfactual statement. To make the matters more concrete, consider the following counterfactual \cite{fine1975critical} 
(3) If Nixon had pressed the button, then there would have been a nuclear holocaust. A similarity-based approach requires the following truth-evaluation: (3) is true if and only if the worlds most similar to the actual world in which Nixon pressed the button, there was a nuclear holocaust. But the worlds in which there is a nuclear holocaust are drastically different from the actual world: the entire future history of humanity would be different in such a world. This example points to the difficulties we face in making judgments about the ordering of possible worlds.


The causal modeling approach for interpreting counterfactuals builds on Lewis's ordering of similar worlds. However, it appeals to the cognitive architecture of the human mind in order to resolve the arbitrariness of assumptions about the ordering of the counterfactual worlds. Pearl \cite{pearl2018book} argues that to make sense of the notion of "similarity" we should rely on the fact that we experience the same world and share the same mental model of its causal structure. However, relying on a largely speculative psychological theory of how the human mind handles the infinity of possible counterfactual worlds does not resolve the normative and ethical implications of choosing which possible worlds are the most similar to the actual world. 

Indeed, different epistemic view points might suggest different ordering of possible worlds. After all, humans differ extensively in the standpoints from which they observe the world, and these standpoints influence the formation of causal mental models \cite{harding2004feminist}. From an abstract point of view, a causal model is specified according to a set of nodes, edges, and assumptions. What these nodes and edges represent and how they are interpreted suggest a particular standpoint about the organization of world from the view point of the causal model. The crucial point to remember is that no causal model captures absolutely objective relations in the world. Depending on the convenient assumptions for a causal model, $X$ can be counterfactually dependent on $Y$ in one model but not in another \cite{halpern2011actual}. These convenient assumptions specifying the causal model might enforce some false perceptions about the social world (at the risk of being seriously wrong). This suggests that there is always a view from somewhere, as opposed to a more objective and universal ``view from nowhere'' \cite{nagel1989view} from which we can assess whether a counterfactual is assertible.

\section{Results of our analysis}

So far, we have argued that the use of counterfactuals in fair and explainable machine learning is not straightforward, and that there are various trade-offs and value judgments essential to the use of counterfactuals for ethical machine learning. Examination of all these assumptions produces awareness about various trade-offs, value judgments, or potential harms of using or misusing counterfactuals. Therefore, to aptly use counterfactuals requires bringing forth all implicit and unspecified assumptions about the ontology of the categories on which we run counterfactual analysis as well as the epistemic and the interpretational issues pertaining to the evaluation of counterfactuals. Examination of all these assumptions produces awareness about some unexpected potential harms that can result from the laudable goals of fair and explainable machine learning.

In this section, we offer strategies for specifying  and reflecting on the hidden ontological and  epistemological-semantic assumptions through an interdisciplinary conversation. We summarize the results of our study (Table 1) by suggesting a detailed set of tenets to check and reflect upon before applying counterfactuals to fair and explainable machine learning. Following this set of tenets would enable modellers and algorithmic designers to state unspecified and implicit assumptions about social ontology as explicitly as possible. It also suggests a path to researchers for seeking a variety of justifications in seeing the social world through a counterfactual lens, and to become aware of some potential harms and disadvantages of making sense of fairness and explanations counterfactually. Our results are a necessary step to perform before designing and applying some putative counterfactually fair or explainable algorithms to social contexts.

\begin{table*}
\begin{tabular}{p{3.5cm}p{5cm}p{5cm}}
\toprule
\sc Assumption
& \sc Question
& \sc Example
\\\midrule
Ontological perspective 
& What are the social categories?    
& What is race (or gender)?
\\\midrule
Ontological choice
& What ontological perspective do we choose to adopt, and why? 
& Among different views, what do we take race to be? Social constructivism? Geo-biological ancestry conception of race? why? 
\\\midrule
Ontological knowledge &  How do we know about the social categories? & Who do we consult about the conception of race?

\\\midrule
Semantic choice  & Close-enough-possible-worlds or causal modeling? What is the justification?  &
Why do we choose either of these semantic approaches to counterfactually suppose that Nora is not Latina? How is our choice justified?

\\\midrule
Evaluation reliability  & What happens to the truth value of the counterfactuals of interest if we change the semantic approach? How robust is the truth value of the counterfactual when moving from a close-enough-possible-worlds approach to causal modeling? & Is the truth value for ``If Nora had not been Latina, she would not have been denied admission.'' differ when we choose the semantic approach?

\\\midrule
Similarity choice  &  How do we choose what similarity means in this context? & What do we sacrifice by supposing a particular cluster of similar worlds (rather than other possible clusters of similar worlds) in which an individual is the same except for their race?

\\\midrule
Comparison criteria & What are our chosen criteria for comparing the similar worlds of interests to the actual world? Are these criteria socially warranted? & What characterization for comparing similar worlds justifies keeping (almost) everything about a person fixed except for their race? What does this socially mean?

\\\midrule
Idealization & What do we miss by translating social categories into random variables? & What is left out by translating an individual's race to a random variable?

\\\midrule
Context & How do these categories operate in the world? & How does race function in the world? Does this conflict with the assumptions necessary for counterfactual manipulation of race? 

\\\midrule
Ethical and social harm & Does our ontological preference generate harms in relation to social justice (combating structural injustices)? & Does our ontological preference for what race is generate harms in relation to combating racial injustice?

\\\bottomrule
 \end{tabular} 
\caption{Any use of a counterfactually fair or explainable algorithm (in a social context) involves making several ontological, semantic and ethical choices and judgments. These implicit presumptions, choices, and judgments must be made as explicit and obvious as possible.
}
\label{tab:nissenbaum}
\end{table*}

Table 1 has three columns. The first column provides a category of different kinds of presumptions and choices (ontological and epistemological-semantic) which are necessary to examine before designing and applying counterfactually fair and explainable algorithms. The second column provides the set of questions to ask and answer for articulating the implicit set of assumptions in column one as explicit as possible. The third column gives an exemplar of the questions to answer in the context of a particular social problem. 

Recall the counterfactual ``If Nora had not been Latina, she would not have been denied admission.'' 

Here are the ontological assumptions that should become explicit. First, an explicit statement of the ontological perspective the algorithmic system is adopting. To tweak ``being Latina'' the designers of the system need to specify what race (e.g., Latina) from their perspective is. Second, the designers could discover whether a counterfactual approach inadvertently commits them to a problematic social ontology. They could provide morally and politically appropriate justifications for why, among other options, they choose and adopt this ontological perspective about race. Is it because this conception of race is compatible with some simplistic assumptions about social ontology that are required to use a causal modeling approach? What are the genuine reasons for this choice, in relation to respecting intellectual humility for what we know about race from other disciplines? Third, the assumptions about ontological knowledge should become explicit. For instance, who do we consult about a theory of race, and why?

The epistemological-semantic presumptions and choices that must be made explicit are as follows. First, what is the semantic choice? Will we choose a close-enough-possible-worlds or causal modeling approach? What is the justification for this choice? Does our choice make a difference to the truth evaluation of the counterfactual for this particular context of employment such as Nora not being Latina? Second, how do we account for the evaluation of reliability? How robust is the truth value of the counterfactual when moving from a close-enough-possible-worlds approach to causal modeling? For instance, is the truth value of ``If Nora had not been Latina, she would not have been denied admission.'' differ when we choose either of the semantic approaches? Third, how do we decide about the meaning of similarity in the particular context of employment? What do we sacrifice by supposing a particular cluster of similar worlds (rather than other possible cluster of similar worlds) in which an individual is the same except for their race? Or if we are specifying that everything that is not causally dependent on the tweaked category should remain constant, how do we know what is not causally dependent? Fourth, what are our chosen criteria for comparing the possible similar worlds of interest to the actual world? Are these criteria socially warranted? For instance, what characterization for comparing similar worlds justifies keeping almost everything about a person fixed except for their race? What does this socially mean? Fifth, there are questions about the translation of social categories such as race into random variables that can be appropriately treated by an algorithm, if the semantic choice is causal modeling. What is left out if we translate the conception of race into random variables? Does that matter? Why or why not? Sixth, there are questions about the choice of context. How do social categories (such as race) operate in the world? Does this conflict with the required assumptions for counterfactual manipulation of race? Finally, there are questions about some ethical harms that can result from the use of counterfactual analysis. Does our ontological preference generate harms in relation to some desired social justice agenda? For example, does our ontological preference for what race is generate harms in relation to some affirmative action plans for combating racial injustice?

In sum, Table 1 shows that any trace of a counterfactually fair or explainable algorithm (in a social context) involves making several choices and presumptions. By following these tenets, computer scientists can discuss the validity and the implications of these choices in accordance with other disciplines such as philosophy, social sciences,  and anthropology. To that end, the implicit presumptions and choices will be made as explicit and obvious as possible, and an interdisciplinary conversation can result in concluding whether the counterfactuals should be used in the generation of explanations and fairness in machine learning practice.

\section{Conclusion}
 
Counterfactuals are increasingly applied in machine learning, for example in designing fair and explainable algorithms. This paper provides a detailed set of principles, according to philosophical and social scientific insights, for articulating the implicit and unspecified contextual presumptions and choices made in counterfactual applications. Regardless of which evaluation approach to counterfactuals one takes, this set of principles could help researchers to conduct interdisciplinary conversations and become aware of the potential harms and ethical impacts of their counterfactual thinking as it pertains to the social world. We think this set of principles is an example of how to establish a successful interdisciplinary conversation between machine learning researchers and social scientists, philosophers, and ethicists.

\section{Acknowledgements}
We would like to thank Alex Beutel, Yoni Halpern, Manasi Joshi, Christina Greer, Robert Williamson, Mario Günther, and members of the Humanizing Intelligence Grand Challenge at Australian National University for extremely helpful comments and feedback. We would also like to thank participants in the Workshop on Philosophy and Medical AI at the University of Tübingen, NeurIPS's Workshop on Algorithmic Fairness through the Lens of Causality and Interpretability, and the Bias and Fairness in AI Workshop in Ghent, Belgium for critical discussion.

\bibliographystyle{ACM-Reference-Format}
\bibliography{ref}

\end{document}